# Eye Tracking to Understand Impact of Aging on Mobile Phone Applications

Antony William Joseph[1], Jeevitha Shree DV[2], Kama Preet Singh Saluja[3], Abhishek Mukhopadhyay[4], Ramaswami Murugesh[5] and Pradipta Biswas[6]

[1] IT-Integrated Design, National Institute of Design, Bengaluru, Karnataka, India
`antony.william@gmail.com`
[2] I3D Lab, CPDM, Indian Institute of Science, Bangalore, Karnataka, India
`jeevithas@iisc.ac.in`
[3] I3D Lab, CPDM, Indian Institute of Science, Bangalore, Karnataka, India
`kamalpreets@iisc.ac.in`
[4] I3D Lab, CPDM, Indian Institute of Science, Bangalore, Karnataka, India
`abhishekmukh@iisc.ac.in`
[5] Department of Computer Application, Madurai Kamaraj University, Madurai, India
`mrswami123@gmail.com`
[6] I3D Lab, CPDM, Indian Institute of Science, Bangalore, Karnataka, India
`pradipta@iisc.ac.in`

**Abstract.** Usage of smartphones and tablets have been increasing rapidly with multi-touch interaction and powerful configurations. Performing tasks on mobile phones become more complex as people age, thereby increasing their cognitive workload. In this context, we conducted an eye tracking study with 50 participants between the age of 20 to 60 years and above, living in Bangalore, India. This paper focuses on visual nature of interaction with mobile user interfaces. The study aims to investigate how aging affects user experience on mobile phones while performing complex tasks, and estimate cognitive workload using eye tracking metrics. The study consisted of five tasks that were performed on an android mobile phone under naturalistic scenarios using eye tracking glasses. We recorded ocular parameters like fixation rate, saccadic rate, average fixation duration, maximum fixation duration and standard deviation of pupil dilation for left and right eyes respectively for each participant. Results from our study show that aging has a bigger effect on performance of using mobile phones irrespective of any complex task given to them. We noted that, participants aged between 50 to 60+ years had difficulties in completing tasks and showed increased cognitive workload. They took longer fixation duration to complete tasks which involved copy-paste operations. Further, we identifed design implications and provided design recommendations for designers and manufacturers.

**Keywords:** Ocular Parameters . Aging . Cognitive Workload . Task Performance . Mobile Interaction . User Behaviour . User Experience . User Interface

# 1 Introduction

Population aging is an important emerging demographic phenomenon in the world today. Due to advancement in public health and medicine, life expectancy has improved to a great extent. The 2001 census of India has shown that elderly population in India accounted for 77 million and number will rise to nearly 140 million by 2021 [1]. As society is progressively moving towards digital technology, aging population do not want to exclude themselves. We may notice that social support system is going through a transformation, and mobile phones have become an enabling factor to remain connected with family and friends. Introduction of smart technology has impacted aging population [2] in terms of accessibility and utilization. In context of aging population and technology adaptation, it is essential to understand aging user behavior on mobile phones.

Visual sense is an important factor to explore the world around us by moving the eyes. Eye movements are vital information for understanding cognitive process. They uncover cognitive activities and indicate the target of our visual interest [3]. In recent years, eye tracking technology has become a promising tool [4] in Human Computer Interaction (HCI).

Eye tracking has been used for more than a century in psychology for recording eye movements while reading. As personal computers were flourishing, in 1980's, researchers started applying eye tracking techniques into issues of HCI. Later, due to technological advances, eye tracking methods were employed to answer many usability issues [5]. Eye tracking is the process of measuring either point of gaze or motion of an eye relative to head with a device called Eye tracker. The technique gives numerous possibilities to understand users' mental model and problems faced by them while performing tasks. Eye movement is assumed to indicate the thought in relation to cognitive process [6], thereby suggesting to provide dynamic evidence to users' attention in relation to a visual display. In our study, we used a wearable eye tracking glass to analyze aging user behavior while performing simple to complex tasks on a mobile phone. Main contributions of the paper include:

1. Design studies to understand aging user behavior based on essential requirements for fulfilling daily activities on an android mobile phone
2. Investigate how aging affects user experience on mobile phones while performing complex tasks
3. Identify how aging affects a users' cognitive load while performing tasks on a mobile phone.

# 2 Related Work

From literature, we identified few studies that have evaluated effect of age on usability of mobile devices. Rogers et al. [7] conducted a study to evaluate how task demands and users' age influenced task performance on touch and non-touch screen devices. Their study involved 40 younger (18–28 years) and 40 middle-aged to older (51–65 years) participants. They used control tasks such as scrolling, up/down buttons, list boxes, and text boxes. They noted that older adults were slower than younger adults on pointing and scrolling tasks on a touch screen. Moreover, they noted that small button sizes were particularly problematic for the older adults.

On the other hand, Sakdulyatham et al. [8] conducted a study to investigate how user interface of LINE application affects elderly when using smartphones. Their study involved 38 elderly participants (60-69 years) who performed 8 different tasks on a mobile phone. Results showed that due to decline in physical health, elderly group had difficulty in viewing content on screen leading to more errors. The study noted the user interface, with Arial Unicode MS font with size 12 to 16 pts and screen brightness at 75% is an appropriate integration for the elderly.

Moumane, K. et al. [9] conducted an empirical study to evaluate influence of screen size of mobile devices for stability of apps using ISO 25062 and ISO 9241 standards for objective measures. They used Recordable.Mobi for eye tracking. Their study identified a set of challenges when using apps such as screen size, display resolution and capacity of memory. The study suggested that a smartphone with large screen size, facilitated ease of use.

Majrashi et al. [10] conducted an exploratory study to investigate relationship between eye tracking metrics and cross-platform usability problems, using an online shopping application. Their study involved 31 participants aged between 18-60 years. Results from the study noted that, usability issues were associated with prior experience and knowledge.

Al-Showarah et al. [11] conducted a study to examine eye movements for young and elderly participants, to find dissimilarities in browsing on different smartphone and tablet applications. Their results found that elderly participants have high dissimilarity than younger ages. In other words, elderly participants were less efficient in browsing smartphone applications than younger participants.

Al-Showarah et al [12] conducted another study to investigate effect of age on smartphones using eye tracking technology. They evaluated usability of smartphone interfaces for three different age groups: elderly(60+ years), middle age (40 to 59 years) and younger age (20 to 39 years). They extracted eye metrics like fixation duration, scan-path duration and saccades amplitude for their analysis. The authors noted that, elderly participants were less efficient having lower cognitive ability in browsing smartphone interfaces. They found that elderly participants exhibited greater difficulties in processing information on smartphones across all screen sizes than users of middle and younger age groups. In general, they noted that, there exists a possibility of positive relationship between getting older and less experience in using smartphones.

In summary, we may note that number of studies measuring cognitive workload are often limited to a smaller group or only elderly age groups, and not adequately focused across different age groups at large. Thus, we aim to investigate how aging in users of different age groups can affect their experiences on using mobile applications while they perform simple to complex tasks.

## 3    Research Methodology

This study was designed to understand aging user behavior based on essential requirements for fulfilling daily activities on a mobile phone.

**Participants:** Initially 180 participants showed interest for the study, but we could select only 50 participants due to various sociocultural factors and selection criteria followed to recruit participants. Short Portable Mental Status Questionnaire (SPMSQ) was used to select and assess mental status of participants [13]. Participants with score 8 and above out of 10 were selected. In addition to considering SPMSQ, eye information was collected to identify participants who are comfortable with near vision and could read mobile screen with eye tracking glasses without any difficulty. Participants were grouped into five different age groups: Group-A (20-29 years), Group-B (30-39 years), Group-C (40-49 years), Group-D (50-59 years) and Group-E (60 year and above). Each group consisted of 10 participants (5 male and 5 female), the oldest being 77 years of age. However, other inclusion criteria considered were literacy, usage of android phone and residing in Bengaluru City.

**Material:** Samsung Galaxy S7 Edge smart phone was used, with octa core (2.3 GHz, Quad core, M1 Mongoose + 1.6 GHz, Quad core, Cortex A53) processor. It runs on Samsung Exynos 8 Octa 8890 Chipset with 4 GB RAM and 32 GB internal storage. It has 5.5-inch screen and 2560x1440 resolution with 535 pixels per inch. For eye tracking, Tobii Pro Glasses 2 with a sampling rate of 100Hz was used. Tobii Pro Lab software was used to extract raw data and analyze visual information.

**Design:** We used a non-invasive wearable eye tracking glasses. The language used to communicate with participants was English and they were approached in their residence with prior appointment. Bengaluru city was chosen for our study because of its multi-cultural nature and home to large number of people migrated from other Indian states. Prior to designing the five tasks, we conducted a survey on technology prior experience; commonly used mobile applications; usage of mobile phones; frequency of usage of various digital technologies; familiarity and awareness of mobile control buttons in general for all 50 participants. Thus, by considering results of above-mentioned survey, we designed five tasks which are essential for fulfilling daily activities on a mobile phone. For each task, scenario and description were given to contextualize the experiment. It may be noted that, participants who failed to complete given five tasks were also considered in the study.

Further, each participant was asked to perform a set of five tasks on the mobile phone. The five tasks are- Task-1: Adding a phone number to contact; Task-2: Sending birthday SMS greeting; Task-3: Google search and save information in Memo App; Task-4: Online shopping using Amazon app and Task-5: Sending WhatsApp Message. The first two tasks were simple and following three tasks were complex. This was designed to motivate participants and to understand user experience and usability aspect of mobile applications.

**Procedure:** Eye tracking system was calibrated for every participant before beginning tasks. An observer was present throughout the experiment. We noted that most participants were unable to complete at least one task due to complex and counterintuitive mobile haptic interface. Among participants of Group-A, 1 participant failed to complete Task-5; in Group-B, 1 participant failed to complete Task-2 and 4; 2 participants each failed to complete Task-3 and Task-4 in Group-C; in Group-D, 6, 4 and 1 participants failed to complete Task-3, 4 and 5 respectively; and among Group-E users, 8 and 6 participants failed to complete Task-3 and 4 respectively. We found that Task-3 was most difficult as many participants were unable to complete the same. Participants of Group-E were unable to complete a few tasks when compared to other user groups. This may be due to lack of precision of finger movements to perform task on a touch screen.

## 4 Results

We recorded ocular parameters like fixation count, saccadic count, average fixation duration, maximum fixation duration and standard deviation of pupil dilation for left and right eyes for each participant. By analyzing these parameters (also called variables), we aim to identify if aging affects user experience on mobile phones while performing complex tasks, and estimate their cognitive workload while performing different tasks on a mobile phone. We extracted ocular parameters like recording timestamp (in milliseconds), pupil diameter left (in millimeter), pupil diameter right (in millimeter), eye movement type (fixation, saccade, eyes not found, unknown eye movement) and eye movement type index (number) from eye tracking glasses. Number of gaze fixations & saccades were extracted using IV-T fixation [14] filtering algorithm of Tobii Studio software with velocity-based threshold set at 30°/sec.

We observed that average time taken to complete five tasks increased as age of participant increased. Group-E aged over 60+ years took longest time (161.09 seconds) to complete five tasks when compared to participants of other age groups. Participants of all age groups took longest time to complete Task-5 (203.94 seconds).

We conducted ANOVA to identify the main effect and interaction effect of different ocular parameters for participants of different age groups while they performed five tasks. Assumption criteria like independence of samples, experimental errors of data are normally distributed, equality of variance were met, before we conducted ANOVA test. We considered Fixation Rate (FC), Saccade Rate (SC) Maximum Fixation Duration (MFD), Average Fixation Duration (AFD), Standard Deviation Pupil Dilation Left (SPL) and Standard Deviation Pupil Dilation Right (SPR) as the six dependent variables. Here, FC and SC are nothing but fixation count per second and saccade count per second respectively. We found that, main effect of task on FC, SC, AFD, SPL and SPR was not significant ($p>0.05$). On the other hand, main effect of age group for FC, SC, AFD, MFD, SPL and SPR was found to be significant with $p<0.05$. This indicated that aging has a positive effect with task performance on mobile phones.

Additionally, we noted that there was no interaction effect between task and age group on FC, SC, AFD, SPL and SPR, since $p>0.05$. But we noted a significance ($p<0.05$) in the main effect of both task and age group on maximum fixation duration ($F(16,4)=5.203, P<0.001, \eta^2=0.085$; $F(16,4)=6.364, P<0.00, \eta^2=0.102$) and interaction effect between task and age group ($F(16,4)=2.291, P<0.05, \eta^2=0.140$). This indicates that irrespective of age, most participants took longer time to pay attention in completing the five tasks.

| Task |  |  |  |  |  |
|---|---|---|---|---|---|
|  | T1 | T2 | T3 | T4 | T5 |
| T1 |  |  | MFD | MFD | MFD |
| T2 |  |  | MFD | MFD | MFD |
| T3 | MFD | MFD |  |  | MFD |
| T4 | MFD | MFD |  |  |  |
| T5 | MFD | MFD | MFD |  |  |

| Group |  |  |  |  |  |
|---|---|---|---|---|---|
|  | A | B | C | D | E |
| A |  | MFD | MFD | FC, SC, AFD, MFD, SPL, SPR | FC, SC, AFD, SPL, SPR |
| B | MFD |  |  | FC, SC, AFD, MFD, SPL, SPR | FC, SC, AFD, SPL, SPR |
| C | MFD |  |  | FC, SC, MFD, SPL, SPR | SPL, SPR |
| D | FC, SC, AFD, MFD, SPL, SPR | FC, SC, AFD, MFD, SPL, SPR | FC, SC, AFD, MFD, SPL, SPR |  | AFD, MFD |
| E | FC, SC, AFD, SPL, SPR | FC, SC, AFD, SPL, SPR | FD, SPL, SPR | AFD, MFD |  |

**Fig. 1.** Ocular parameters that were found significant with LSD test for Task-wise (top) and Group-wise (bottom)

We considered post hoc comparison using LSD test for all ocular parameters and noted statistically significant difference between two means for within task and within age group elements. Based on number of ocular parameters that were found to be significantly different, we assigned color coding scheme as shown in Fig. 1. We found more green and brown areas for within age group when compared to within tasks. This indicated that most ocular parameters were found to be significantly different for within age group elements. Ocular parameters FC, AFD, SPL and SPR were found to be significantly different for within age group elements but not for within task elements. This indicates that age has bigger effect on performance irrespective of task the participants performed. Fig. 1 shows color-coding scheme for indicating significance of ocular parameters considered for our study. In Fig.1, each color is allocated to each of the boxes based on number of variables significant. Color Green is assigned when a single variable out of six variables is significant, color Orange is assigned when two out of six variables are significant, color Brown when three out of six variables are significant, color Pink when five out of six variables are significant and color blue when all six variables are significant. Color Yellow is assigned when all six variables are not significant.

Further, we performed fixation analysis for each task (Fig. 2) and age group (Fig. 3) respectively. Results state that Task-5 recorded higher fixation rate (Fig. 2) requiring longer duration of visual search. Task-3 where participant had to Google search and save information in Memo App took longest fixation duration when compared to other tasks indicating that participants took longest time to pay attention to complete task. Participants of Group-D and Group-E had higher rate of fixation (Fig. 2) when compared to participants of other age groups. This means that the above two groups had difficulty in performing and completing five tasks given to them. We further noted that, Group-B and Group-C had longer fixation duration values when compared to the rest.

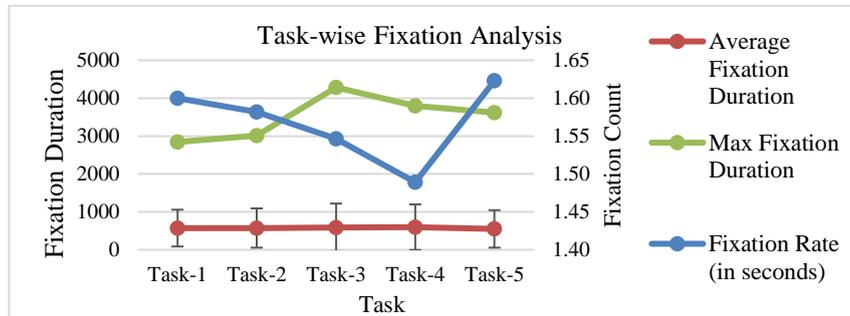

**Fig. 2.** Task-wise fixation analysis.

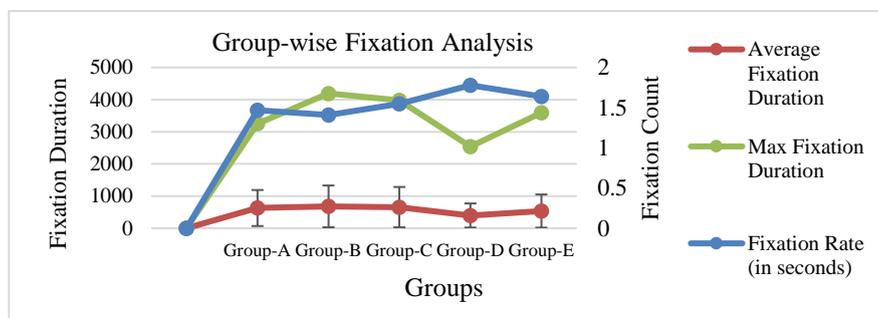

**Fig. 3.** Group-wise fixation analysis.

From Fig. 2, we may note that, average time taken to complete task is increasing linearly from Task-1 to Task-5 respectively and participants of all age groups took longest time to complete Task-5 (203.94 seconds) when compared to other tasks. We noted a similar trend in age group wise analysis. Fig. 4 and Fig. 5 indicate pupil dilation of tasks and age groups respectively. We noted that standard deviation of pupil dilation was high while participants performed Task-3 and Task-4. Participants of age Group-D and Group-E showed higher values of standard pupil dilation.

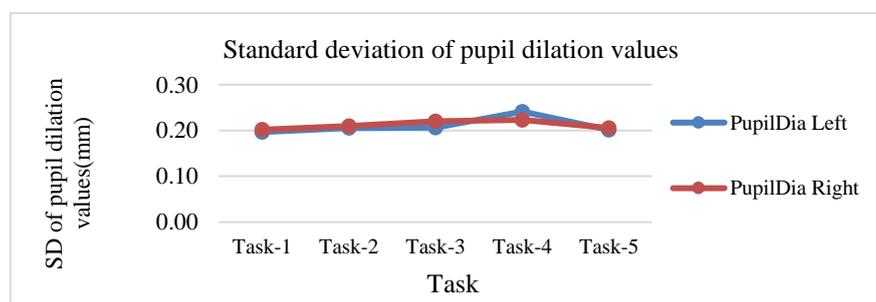

**Fig. 4.** Standard deviation of pupil dilation for five tasks.

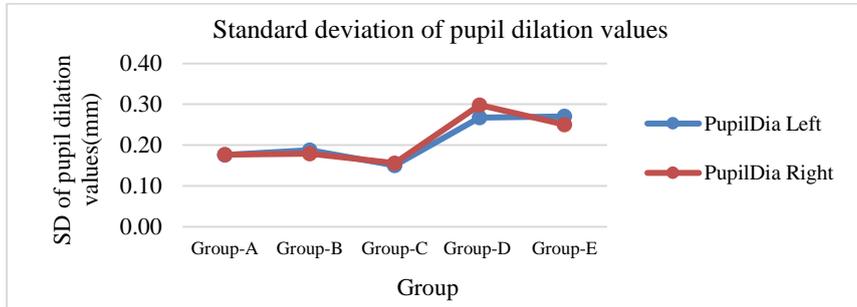

**Fig. 5.** Standard deviation of pupil dilation for five tasks.

## 5       Estimating Cognitive Workload

We identified how aging affects users' cognitive workload of each participant in two ways. First, we analyzed ocular parameters like fixation frequency, fixation rate, saccade rate, fixation duration and standard deviation of pupil dilation for left and right eye values. Study from Coral [15] noted that by analyzing above ocular parameters, we can estimate cognitive workload of participants of different age groups for different tasks. Coral [15] presented a summary of eye-related measurements and their relationship to increased cognitive workload. From summary, it can be noted that increase in fixation frequency, fixation rate, saccade rate, fixation duration and pupil dilation values indicate increase in cognitive workload. From above results, we can see that participants of Group-D and Group-E were estimated to have higher cognitive workload when compared to rest as they showed increased values in above-mentioned ocular parameters. All participants showed highest estimation of cognitive workload while performing Task-3 and Task-4.

Further, as a second way to estimate cognitive workload for each participant, we analyzed pupil dilation of both left and right eyes for each participant. We developed an algorithm based on pupil dilation using low pass filter (LPF). We divided pupil diameter signal into sections of 100 samples (time buffer of 1 second as sampling frequency of device is 100Hz). We removed DC offset from original signal by subtracting the mean of original signal. Then, we have applied Butterworth lowpass filter to filter signal within 0 to 5Hz and summed up magnitude of filtered signal in a running window of 1 second with 70% overlap between adjoining sections. The algorithm is described in Fig. 6.

Each participants' pupil diameter values were given as input to our algorithm. The algorithm gives a single value as output which is nothing but average LPF for both eyes which can be used to estimate cognitive workload. This procedure is performed for both left and right pupil diameter values for participants of all age groups and for all tasks. Fig. 7 and Fig. 8 show average LPF for left and right eyes for group and task respectively. This can be used to estimate average cognitive workload for participants of all age groups while performing all tasks. It is evident that, participants of Group-D and Group-E were estimated to have higher cognitive workload (higher average LPF) when compared to the rest. We can see from graph (Fig. 8) that, Task-3 was estimated to have higher cognitive workload (higher average LPF).

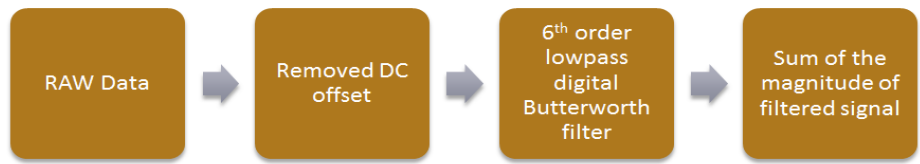

**Fig. 6.** Algorithm to estimate cognitive workload

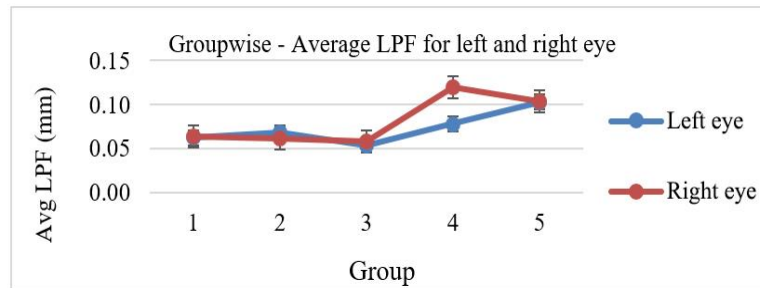

**Fig. 7.** Group-wise average LPF

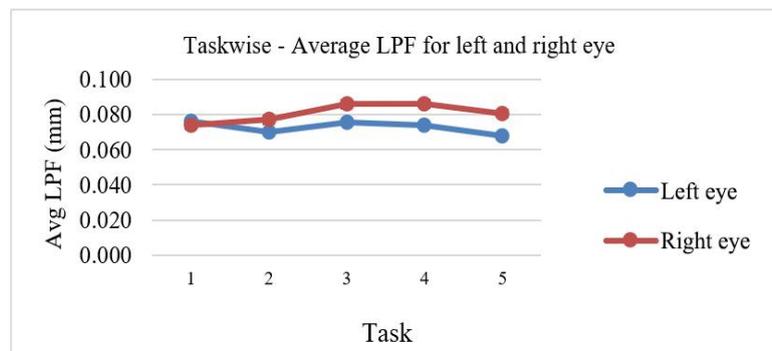

**Fig. 8.** Task-wise average LPF

## 6  Discussion

Taken together, above statistical results show that all ocular parameters considered were found to be significant across all age groups. This indicates that aging affects user experiences on mobile phones while performing complex tasks. Participants of Group-D and Group-E aged between 50-59 years and 60+ years took longer time to complete each task and showed to have higher cognitive workload while performing tasks. Task-1 and Task-2 were perceived as easy task and Task-3 was perceived as the most difficult task of all. Additionally, our results noted that participants of all age groups took longest time to complete Task-3. It was found that 2 participants from Group-C, 6 participants from Group-D and 8 participants from Group-E were unable to complete the same. Task-3 showed maximum fixation duration indicating that participants possessed greater attention to complete task.  This was due to task involving copying a text from browser and saving the same in Memo App.  Though copy-paste operation is familiar to participants, they struggled due to touchscreen interface that was not intuitive. They have problem with soft keys on full touchscreen phones and fail to map physical keys and soft keys. For example, Lee et al. [16] reported that minimal target size for tapping with good speed and accuracy appears to be 8 mm for all age groups. Another important issue discovered is that older participants could not differentiate between short press and long press on touch screen.  As a result, controls were not released on time and options distracted their attention. It was noticed that a few participants from Group-C and Group-D tried to pinch and zoom page for selecting and copying text but failed due to unavailability of function. Task-4 was recorded as second highest fixation duration and was perceived as second difficult task of all. It was noticed that 2 participants from Group-C, 3 participants from Group-D, 6 participants from Group-E were unable to complete the same. This was due to poor representation of interface elements and usability issues that did not serve natural user behavior like 'product search', 'add to cart' and 'shopping cart icon'. Many pressed 'shopping cart icon' on right top corner of application instead of clicking the 'add to cart button' for adding selected product but failed in completion of task. Therefore, we can infer that there is a positive correlation on how aging affects users' cognitive workload while performing complex tasks on a mobile phone. However, there is no significant difference in interaction effect between task and group.

Participants were able to comprehend all tasks well but struggled to achieve their goal due to touchscreen-based interface. One major problem identified was usability issue of ineffective representation of visual elements for user interaction. There was a mismatch between user interface elements and users' mental model which led to poor performance of tasks and increased cognitive workload. Task performance was correlated with different age groups, time taken to complete task and number of tasks completed successfully. As people age, they change in numerous ways both biologically and psychologically. Critical cognitive functions most affected by age are attention, memory, speed of processing and problem solving. In this context, above results provide facts that aging affects users' cognitive workload while performing tasks on smartphone. Mobile interface was perceived as complex due to usability issues in terms of finding right option to complete task and achieve goal. It is found that familiarity with a feature allows user to use interface quickly and intuitively. Unfamiliar words, symbols and less well-known functions add complexity to task and increases cognitive load. Lack of clarity and ambiguity observed throughout Tasks 3 to 5 led participants to exhibit a sense of frustration and anxiety. Czaja et al. [17] said that effect of technology use needs to be addressed by developers to accommodate all potential users. Though touch screen interfaces are typically designed to be operated by users' hand, alternate designs appropriate for tactile and motor systems are required to be seriously investigated for elderly users.

Based on our eye tracking experiment study findings, we propose few design guidelines for designers and developers to design intuitive and adaptable interfaces for different age groups. Involving elderly people in early design process and investigating possible steps of users' path will enable designers to address interaction issues that may occur. One good approach to designing adaptable intuitive user interface is applying familiar metaphors to user interface elements. Basically, the design needs to represent users' mental model which is highly recommended. For example, in Amazon online shopping application, elderly users repeatedly pressed the 'shopping cart icon' for adding products but in reality, user has to press the 'add to cart' button which is placed just below the product description. Elderly users typically make mistakes when interacting with small targets on mobile screen. In our study, we noticed that touch area was very small and older users struggled to perform actions. Thus, interactive objects should be larger than 8mm, facilitating better user experience and this agrees with previous research [11]. Another important issue which needs greater attention is that, the elderly users find difficult to differentiate between short press and long press on touch screen and they often miss intended targets due to large size of their fingers. Providing intuitive interaction through natural-language interface would benefit them.

## 7    Conclusion

This study aims to investigate how aging affects user experience on mobile phones while performing simple to complex tasks and identify how aging affects a users' cognitive workload while performing complex tasks on a mobile phone. We designed an experiment and collected data from 50 participants aging from 20 to 60+ years. Participants were grouped into five different groups based on their age. Five tasks were designed, and each participant was asked to undertake all the five tasks. From results, we conclude that, as users age, their experience on using mobile phones to perform everyday activities reduces. This can lead to having higher cognitive workload as they age and when the complexity of task is increased. We observed a mismatch between user interface elements and users' mental model which led to poor performance of tasks and increased cognitive workload. As a part of future work, we can develop a prediction model which can accurately estimate cognitive workload of each participant while performing various tasks on a mobile phone and help designers improve mobile user interface.

**Acknowledgment:** The authors are thankful to the Department of Computer Application, Madurai Kamaraj University; I3D Laboratory, CPDM, IISc Bangalore and National Institute of Design, Bengaluru Campus for their encouragement, motivation and relentless support in carrying out our study.